\documentclass{article}[11pt]

\newcounter{extendedAbstract}
\usepackage{ifthen}

\newcommand{\extendedabstract}[1]{
\ifthenelse{\isodd{\theextendedAbstract}}{#1}{}
}

\newcommand{\fullversion}[1]{
\ifthenelse{\isodd{\theextendedAbstract}}
{}
{#1}
}

\newcommand{\tinytitle}[1]{\vspace{-11pt} \paragraph{#1}}

\usepackage{amsfonts,amsmath}
\usepackage{graphicx}

\usepackage{mathptmx} 

\usepackage[latin1]{inputenc} 
\newtheorem{lemma}{Lemma}
\newtheorem{definition}{Definition}
\newtheorem{theorem}{Theorem}
\newtheorem{corollary}{Corollary}

\newcommand{\sgn}{\mathrm{sgn}}

\newcommand{\RR}{\mathbb{R}}

\newcommand{\NN}{\mathbb{N}}

\newcommand{\cost}{\text{cost}}
\newcommand{\OPT}{\text{OPT}}

\newcommand{\depth}{\text{depth}}

\newcommand{\Q}{Q}

\let\margin\marginpar
\newcommand\myMargin[1]{\margin{\raggedleft\scriptsize\textcolor{black}{#1}}}
\renewcommand{\marginpar}[1]{\myMargin{#1}}

\usepackage{color}

\begin{document}

\title{Simpler Proofs by Symbolic Perturbation
}

\author{
Tobias Jacobs\thanks{This work was supported by a fellowship within the Postdoc-Programme of the German Academic Exchange Service (DAAD).}\\
\normalsize
National Institute of Informatics\\
\normalsize
2-1-2 Hitotsubashi, Chiyoda-ku, Tokyo 101-8430.\\
\normalsize
 \texttt{jacobs@nii.ac.jp}. 
}
\date{}
\begin{titlepage}

\maketitle

\thispagestyle{empty}

\begin{abstract}

In analyses of  algorithms, a substantial amount of effort  has often to be spent on the discussion of special cases. For example, when the analysis considers the cases $X<Y$ and $X>Y$ separately, one might have to be especially careful about what happens when $X=Y$. On the other hand,  experience tells us that when a yet unregarded special case of this kind is discovered, one nearly always finds a way to handle it. This is typically done by modifying the analysis and/or the algorithm very slightly.
	
	In this article we substantiate this observation theoretically. We concentrate on deterministic algorithms for weighted combinatorial optimization problems. A problem instance of this kind is defined by its structure and a vector of weights. The concept of a \emph{null case} is introduced as set of problem instances whose weight vectors constitute a nowhere open set (or null set) in the space of all possible weight configurations. An algorithm is called \emph{robust}  if	any null case can be disregarded in the analysis of both its solution quality and resource requirements. 
	
We show that achieving robustness is only a matter of breaking ties the right way. More specifically, we show that the concept of \emph{symbolic perturbation} known from the area of geometric algorithms guarantees that no surprises will happen in null cases. We argue that for a huge class of combinatorial optimization algorithms it is easy to verify that they implicitly use symbolic perturbation for breaking ties and thus can be analyzed under the assumption that some arbitrary null case never occurs. Finally, we prove that there exists a symbolic perturbation tie breaking policy for any algorithm.
\end{abstract}

\end{titlepage}

\section{Introduction}
\label{sec:introduction}

Let $f:\RR^n \to \RR$ be a continuous function and let $C \subset \RR^n$ be a null set. It is a well-known fact that $\sup\{f(x) \mid x \in \RR^n \setminus C \} = \sup\{f(x) \mid x \in \RR^n\}$. 

Imagine that $A(x)$, the strictly positive value of the solution computed by an algorithm $A$ that takes vectors of $\RR^n$ as the input, is continuous. Assume that the value of the optimal solution $OPT(x)$ is as well continuous. Let there be a proof that $A$ is a $c$-approximation which holds for all almost every input, i.e. $OPT(x)/A(x) \leq c$ for each $x \in \RR^n \setminus C$, where $C$ is a null set. Then, as $OPT/A$ is continuous as well, we know that $A$ is a $c$-approximation also for inputs from $C$.

$C$ is what we refer to with the term \emph{null case}. If in the  above scenario it is known a priori that $A$ and $OPT$ are continuous, one can conveniently disregard some arbitrary null case in the proof for $OPT(x)/A(x) \leq c$. For example, one could assume that $x_i \neq 0$ and $x_i \neq x_j$ for each $i \neq j$. We say that $A$ is \emph{robust}.

\tinytitle{Motivation.} This work is motivated by the desire to banish the nasty corner cases that deface otherwise beautiful proofs so often. When some proof distinguishes between cases like $X<Y$ and $X>Y$, one often has to be very careful to which case to assign the special situation $X=Y$, and sometimes this event even has to be analyzed as a third case. One type of equality is especially annoying: when there is more than one optimal solution to a combinatorial problem. A common experience is that one wants to prove a certain property of the optimal solution, but at some point one realizes that it can only be shown that there exists one optimal solution having the property. So much more elegant proofs could be - in a world without equality! 

Inspired by the observation that degenerate cases of the kind just described are null sets in the space of weight configurations, we seek to investigate to what extent the above continuity argument can be exploited in the area of combinatorial optimization. 

The main obstacle is that one cannot assume that $A(x)$, the value of the solution computed by algorithm $A$, is continuous. Consider for example the problem PARTITION, where $n$ numbers $x_1,\ldots,x_n$ have to be assigned to two sets $S_1,S_2$ such that 
 \begin{equation}
 \label{eq:partition}
 \sum_{x_i \in S_1} x_i- \sum_{x_i \in S_2} x_i
 \end{equation}
 is minimized. Approximation algorithms for this NP-hard optimization problem usually have some points of discontinuity. These points lie on the border where the algorithm switches its output from one partition to to another. In contrast, the value of the optimal solution OPT is continuous, because at the points where OPT switches its output, say from $(S_1,S_2)$ to $(S_1^\prime,S_2^\prime)$, the cost of $(S_1,S_2)$ equals the cost of $(S_1^\prime,S_2^\prime)$. However, this observation by itself does not allow to disregard some null case while proving that $A$ is optimal, because the continuity of $A$ must be known a priori.

\tinytitle{Our contribution.}
We identify a sufficient precondition for robustness that is much weaker than continuity. As the starting point serves the observation that the source of discontinuity in algorithms is conditional branching, as there is no other way to compute non-continuous functions in common machine models. Our point of view on algorithms is the decision tree model. In this model an algorithm branches by comparing the value $v(w)$ of some continuous \emph{branching function} $v: \Q^n \to \Q$ with 0, where $w$ is the weight vector of the problem instance.
 The algorithm takes one branch if $v(w) < 0$ and the other if $v(w) > 0$; ties are broken in some well-defined manner. 
 
It will become clear that in this model the points of discontinuity are exactly (a subset of) the cases where there is a tie that has to be broken by the algorithm. We show that a generalization of the symbolic perturbation technique proposed in the context of geometric algorithms~\cite{Neuhauser97,Yap90} (see Section~\ref{sec:summary} for a discussion of related work) guarantees robustness.

We complement our study with a theorem which states that there exist a symbolic perturbation policy for any algorithm. That theorem, although not being constructive (it might be arbitrarily complicated to find a policy), supports our belief that the concept of symbolic perturbation is widely applicable.

We wish to emphasize that this work proposes the employment of symbolic perturbation in \emph{analyses} of algorithms. 
Our goal is to simplify analyses by recognizing the situation where algorithms \emph{implicitly} make use of symbolic perturbation. We believe that for a huge class of algorithms this holds true and is rather simple to verify. Once one has shown that an algorithm employs symbolic perturbation, one can choose an arbitrary null case and assume during the analyses that it never occurs. In other words, we understand symbolic perturbation as a proof technique rather than an algorithm design technique.

\tinytitle{Paper organization.}
In the following section we provide the mathematical basics of our result. The main theorem of that section is a generalized version of the fact pointed out in the first two lines of this article. Then, in Section~\ref{sec:model}, the model for combinatorial problems and algorithms is introduced. In Section~\ref{sec:mainresult} we formally define what a null case is and also introduce our notion of symbolic perturbation. After that, we show that algorithms using symbolic perturbation are robust.  The practicability of this result is demonstrated by a complete example given in Section~\ref{sec:example}, and in Section~\ref{sec:existence} we prove that there exists a symbolic perturbation policy for any algorithm.
Section~\ref{sec:summary} summarizes the findings and discusses directions for future research. Here we also compare our approach to other areas where symbolic perturbation has been proposed.
\extendedabstract{Proofs omitted or sketched in this extended abstract appear in the full version of the paper (see appendix).}

\section{Mathematical background}
\label{sec:math}

In the introduction we have considered an algorithm that takes vectors from $\RR^n$ as the input, and a null case has been a null set in that space. However, we want to avoid that the results are only shown for the rather abstract computational model where weights can be any rational number.
We therefore assume that the numbers operated with belong to a subset $\Q \subseteq \RR$. Depending on the model of computation, $\Q$ can be equal to $\RR$, $Q$ can be the set of rational numbers, or the set of algebraic numbers. Although not necessary, it makes sense to assume that $\Q$ is dense in itself, i.e. for any $x \in \Q$ and $\epsilon>0$ there is some $y \in \Q$ with $0<|x-y|<\epsilon$. The assumption is sensible because isolated points never belong to null cases.

The notion of null sets has been used in the introduction for illustrating the ideas, but in the remainder of the paper no measure theory is employed. It turns out that null cases can as well be defined, in a simpler and even more general way, in terms of subsets of the metric space $(\Q^n,d)$. In principle, our results hold regardless of the metric $d$, but for simplicity we assume that $d$ is the Euclidean distance.

%

The following definitions can be found in any introductory analysis textbook. A \emph{neighborhood} of a point $x \in \Q^n$ is defined as  $\{y \in \Q^n \mid d(x,y) < \epsilon\}$ for some $\epsilon>0$. A subset of $\Q^n$ is defined to be \emph{open} if it contains some neighborhood of each of its elements. A function $f:\Q^n \supseteq X \to \Q$ is said to be continuous at $x \in X$ if for any $\epsilon > 0$ there exists a $\delta > 0$ such that $d(x,y) \leq \delta \Rightarrow |f(x) - f(y)| \leq \epsilon$ for every $y \in X$. The function $f$ is continuous if it is continuous at every point of its domain.
The following definition corresponds to the notion null sets used in the introduction.
\begin{definition}
\label{def:nowhereopen}
A set $X \subset \Q^n$ is nowhere open if it contains no non-empty open subset.
\end{definition}
The correspondency is given by the fact that in the Euclidean space $\RR^n$, with a measure defined in the standard way, any null set is nowhere open. Definition~\ref{def:nowhereopen} is even more general: e.g. the set of irrational numbers is nowhere open in $\RR$ although it is not a null set. An alternative definition would be to say that a set is nowhere open if it has no interior points.

For the observation pointed out in the first two lines of the paper to hold, it is crucial that the domain of $f$ satisfies a certain condition. For example, if the domain of $f$ was a null set itself the claim would not hold. We formulate the requirement as follows.

\begin{definition}
\label{def:semi-open}
A set $X \subseteq \Q^n$ is called \emph{semi-open} if any neighborhood of its members contains a non-empty open subset of $X$.
\end{definition}

Remark that each open set, including the empty set, is semi-open. Intuitively, a semi-open set is the union of an open set with a subset of its boundary. 
We are now ready to translate the proposition from the first two lines of this article into the terms just introduced.

\begin{lemma}
\label{lemma:semi-open}
Given a non-empty semi-open set $X \subseteq \Q^n$, a nowhere open set $C \subset X$, and a continuous function $f:X \to \Q$, it holds that $\sup\{f(x) \mid x \in X \setminus C\} = \sup\{f(x) \mid x \in X\}$.
\end{lemma}

\tinytitle{Proof.}
Consider any point $x \in C$ and any $\epsilon > 0$. From Definition~\ref{def:semi-open} follows that the $\epsilon$-neighborhood of $x$ contains a non-empty open subset $U$. As $C$ is nowhere open, it cannot contain every point of $U$, i.e. there is a point $x_\epsilon \in X \setminus C$ with $d(x,\epsilon) \leq \epsilon$.

As a consequence, there is a sequence $(x_i)_{i \in \NN}$ with $x_i \in X \setminus C$ for all $i \in \NN$ and $\lim_{i \to \infty} x_i = x$. From the continuity of $f$ follows that $\lim_{i \to \infty} f(x_i) = f(x)$. which implies that $f(x) \leq \sup\{f(y) \mid y \in X \setminus C\}$. \hbox{} \hfill{$\Box$}

As outlined in Section~\ref{sec:introduction}, one cannot hope to achieve that the output of algorithms is continuous in the input. We now specify a weaker property than continuity. The subsequent theorem ensures that this property is sufficient for robustness, and in the subsequent sections it will be shown that suitable tie-breaking strategies effectuate that the output of algorithms satisfy it.

\begin{definition}
\label{def:locally_continuous}
Let $X \subseteq \Q^n$ be semi-open. A function $f:X \to \Q$ is called \emph{locally continuous} if for any $x \in X$ there is a semi-open set $M \subseteq X$ with $x \in M$ and $f|M$ is continuous.
\end{definition}

\begin{theorem}
\label{thm:math}
Let $X \subseteq \Q^n$ be semi-open, let $f:X \to \Q$ be locally continuous, and let $C \subset X$ be a nowhere open set. Then $\sup\{f(x) \mid x \in X \setminus C\} = \sup\{f(x) \mid x \in X\}$.
\end{theorem}

\tinytitle{Proof.} 
Each $x \in C$ belongs to a semi-open subset $M \subseteq X$ where $f|M$ is continuous. From Lemma~\ref{lemma:semi-open} follows that $f(x) \leq \sup\{f(y) \mid y \in M \setminus C\} \leq \sup\{y \mid f(y) \in X \setminus C\}$. \hfill{$\Box$}

\section{Weighted problems and resistant algorithms}
\label{sec:model}

In this work we intend to make provable and general statements about algorithms for weighted combinatorial problems. Therefore, we need a precisely defined model that is general enough to capture a large number of actual problems and algorithms. For simplicity we only consider minimization problems here, but all our results can be shown for maximization problems in exactly the same way. For an example of how an actual algorithm for an actual problem fits into the model we refer to Section~\ref{sec:example}.

A \emph{weighted combinatorial optimization problem} $P$ is a set of problem structures. Each problem structure $S \in P$ is characterized by the number $n$ of weights, the semi-open domain $W \subseteq \Q^n$ of possible weight vectors, the set $L$ of feasible solutions, and the evaluation function $\cost:(W,L) \to \RR$. For any fixed $l \in L$, $\cost(\cdot,l)$ is continuous.
A \emph{problem instance} of $P$ is a pair $(S,w)$ with $S = (n,W,L,\cost) \in P$ and $w \in W$. The objective is to determine a solution $l \in L$ that minimizes $\cost(w,l)$.

A \emph{deterministic algorithm} $A$ for some problem $P$ is formally defined as follows: For each structure $S = (n,W,L,\cost) \in P$ there is a corresponding finite binary decision tree $T$ whose leaves are associated with solutions from the set $L \cup \{l_f\}$, where $l_f \notin L$ is a special solution for modeling the failure of $A$ to compute a feasible solution. Each internal node of $T$ is associated with a continuous \emph{branching function} $v:W \to \Q$. Given a problem instance $(S,w)$ as the input, the algorithm traverses the tree $T$ corresponding to $S$ starting at its root. At each encountered internal node $v$ it evaluates $v(w)$. If $v(w) < 0$ the left branch is taken, and if $v(w) > 0$ the algorithm takes the right branch. In the situation of $v(w) = 0$, called  a \emph{tie}, either the left or the right branch is taken. The decision is made by some deterministic \emph{tie breaking policy}. When the algorithm finally reaches a leaf, it returns the solution $l \in L \cup \{l_f\}$ associated with it.

The reader may have noticed that the definitions include two continuity assumptions. First, we assume that the cost function is continuous for any fixed solution. We are convinced that this barely restricts the applicability of our model. In optimization problems as we know them, the cost function is typically very simple (see e.g. Equation~\ref{eq:partition} in Section~\ref{sec:introduction}). A straightforward consequence of the first continuity assumption is that the value of the optimal solution is continuous as well.

\begin{lemma}
	\label{lemma:opt-continuous}
	Let $P$ be a weighted combinatorial optimization problem, and let $\OPT(S,w)$ denote the value of the optimal solution for instance $(S,w)$.  For any fixed structure $S \in P$ it holds that $OPT(S,\cdot)$ is continuous.
\end{lemma}

\tinytitle{Proof.}
	Let $S = (n,W,L,\cost)$. By definition, $OPT(S,w) = \min\{\cost(l,w) \mid l \in L\}$. As the $\min$-operator is continuous, $OPT$ is calculated as a concatenation of continuous functions.
\hfill{$\Box$}

\tinytitle{}
The second continuity assumption concerns the branching functions. Again, we believe that the behavior of most algorithms can be modeled using only continuous branching functions. One could justify the assumption theoretically by arguing that in machine models the only source of discontinuity is conditional branching. 

We remark that the model only captures algorithms that always terminate. Some comments about how infinite loops could be modeled can be found in Section~\ref{sec:summary}.

\section{Null cases and robust algorithms}
\label{sec:mainresult}

In this section we define null cases formally. We then describe the concept of symbolic perturbation and show that this tie breaking manner leads to algorithms that can be analyzed under the assumption that some arbitrary null case does not occur. As mentioned in the introduction, this is already the main result of this paper, because we refrain from proposing ways to compute symbolic perturbation explicitly.

\begin{definition}
\label{def:nullcase} 
Let $P$ be a weighted combinatorial optimization problem. A \emph{null case} $\mathcal{C}$ is a set of instances of $P$ where for any fixed structure $S = (n,W,L,\cost) \in P$ the set $\{ w \in W \mid (S,w) \in \mathcal{C} \}$ is a nowhere open set in $W$.
\end{definition}

We need to introduce some more mathematical concepts. For $x \in \Q$, $\sgn(x)$ is defined as $x/|x|$ if $x \neq 0$ and $\sgn(0)=0$.

\begin{definition}
Let $h:\RR \to \Q^n$ be a function such that $h(0) = 0$ and $h$ is continuous at $0$. Let further $W \subseteq \Q^n$ be semi-open and $w \in W$.

We say that $W$ continues into direction $h$ at $w$ if there is some $\delta > 0$ such that for any $0 < a < \delta$ it holds that some neighborhood of $w + h(a)$ is contained in $W$.

If $W$ continues into direction $h$ at $w$ and $f:\Q^n \to Q$ is continuous, we say that $f$ is increasing (constant, decreasing) into direction $h$ at $r$, if there is some $\delta>0$ such that for each $0<a<\delta$ it holds that $\text{sgn}(f-f(w))$ is constantly $1$ ($0$,$-1$) in some neighborhood of $w + h(a)$.
\end{definition}

Note that the condition of $f$ being 
constant in some neighborhood of $w + h(a)$ is only required explicitly in the case of $f$ being constant into direction $h$. In the other cases it suffices to demand that $f(w + h(a))>0$ ($<0$) for any $0<a<\delta$, the rest follows automatically from the continuity of $f$.

\begin{definition}
\label{def:symbolic_perturbation}
	Let $A$ be an algorithm for problem $P$. A symbolic perturbation tie breaking policy for $A$ is given as follows.
	
For each problem instance $(S,w)$ with $S=(n,W,L,\cost)$ there is a function $h:\RR \to \Q^n$ such that $W$ continues into direction $h$ at $w$ and any branching function in the tree corresponding to $S$ is either decreasing, constant, or increasing into direction $r$ at $x$.
	
	In case of a tie  at node $v$, algorithm $A$ takes the left branch when $v$ is decreasing into direction $r$ at $x$. Otherwise, the right branch is taken.
\end{definition}

For proving that $A$ uses symbolic perturbation, one has to specify a suitable function $h$ for each instance and then show that the algorithm breaks ties according to it. This might sound more involved than it actually is: in most cases, a small number of simple functions work out for all instances.

The following lemma can be considered as the main technical lemma of this section, because it establishes the link between symbolic perturbation and the property of robustness.

\begin{lemma}
\label{lemma:mainlemma}
Let $A$ be an algorithm for problem $P$. If $A$ employs symbolic perturbation, then any node $u$ in a decision tree traversed by $A$ for some structure $S=(n,W,L,\cost) \in P$ has the property that it is encountered by $A$ for a semi-open subset of $W$.
\end{lemma}

\fullversion{
\tinytitle{Proof.}
}
\extendedabstract{
\tinytitle{Proof sketch.}
}
Let $X \subseteq W$ be the set of weight vectors where node $u$ is traversed by $A$, and let $w \in X$. Let  $h$ be the direction function corresponding to instance $(S,w)$ according to Definition~\ref{def:symbolic_perturbation}. We show that $X$ continues into direction $h$ at $w$, which directly implies the lemma.

We use induction over the tree depth of $u$ and therefore consider $u$ to be the tree root in the base case. Here, $X=W$ and the claim holds by definition.
For the induction step, assume that the claims have already been shown for some node $v$, and consider a child node $u$ of $v$. We only consider the case where $u$ is the right child of $v$, because the argumentation for the opposite case is analogous. Let $X$ and $X^\prime$ be the set of weight vectors where $v$ and $u$ is reached by $A$, respectively. Now consider any vector $w \in X^\prime$.
\extendedabstract{
We distinguish between three different cases that can occur: $v(w) > 0$, $v(w) = 0$ and $v$ is constant into direction $h$ at $w$, or $v(w) = 0$ and $v$ is increasing into direction $r$ at $w$. For each of these cases it is shown separately that $X^\prime$ continues into direction $h$ at $w$.
}
\fullversion
{
 We distinguish between three different cases that can occur.

\emph{Case 1:} $v(w) > 0$. Then, due to the continuity of $v$, there is some neighborhood of $w$ that is completely contained in $X^\prime$. The claim follows from the induction hypothesis.

\emph{Case 2:} $v(w) = 0$ and $v$ is constant into direction $h$ at $w$\footnote{Note that this case does not even occur in the setting where $u$ is the left child of $v$.}. By definition, there is a $\delta>0$ such that for each $0<a<\delta$ it holds that $v(U_a)=0$ for some neighborhood of $U_a$ of $a+h(a)$. As these neighborhoods are open sets, each point $y$ from any $U_a$ has a neighborhood where $v$ is constantly $0$, and so $v$ must be constant into any direction at $y$. Consequently, the algorithm takes the branch to $u$ for weight configuration $y$ as well. In other words, the neighborhoods $U_a$ are completely contained in $X^\prime$, which proves that $X^\prime$ continues into direction $h$ at $w$.

\emph{Case 3:} $v(w) = 0$ and $v$ is increasing into direction $r$ at $w$. By definition, this means that there is some $\delta>0$ such that for any $0 < a < \delta$ there is some $\epsilon_a$ with $v(x)>0$ for any $x \in W$ with $d(w+h(a),x)<\epsilon_a$. The induction hypothesis states that there is some $\delta^\prime$ such that for any $0 < a < \delta^\prime$ there is some $\epsilon_a^\prime$ with $x \in X$ for any $x \in W$ with $d(w+h(a),x)<\epsilon_a^\prime$. These two propositions imply that there is some $\delta^{\prime\prime} := \min\{\delta,\delta^\prime\}$ such that for any $0<a<\delta^{\prime\prime}$ there is some $\epsilon_a^{\prime\prime} := \min\{\epsilon_a,\epsilon_a^\prime\}$ with $x \in X^\prime$ for any $x \in W$ with $d(x,w+h(a))<\epsilon_a^{\prime\prime}$. In other words, $v$ continues into direction $h$ at $w$ in $X^\prime$.
}
\hfill{$\Box$}

\fullversion
{
With one more simple lemma, the main theorem follows rather straightforwardly.

\begin{lemma}
\label{lemma:concatenation}
Let $f:\Q^n \to \RR$ be locally continuous and let $g:\RR \to \RR$ be continuous. Then $g \circ f$ is locally continuous.
\end{lemma}

\tinytitle{Proof.} 
Each point $x$ in the domain of $f$ is contained in some semi-open subset $X$ such that $f|X$ is continuous. As $g$ is continuous, so is $(g \circ f)|X$. \hfill{$\Box$}
}
\extendedabstract
{
\addtocounter{lemma}{1}
}

\begin{theorem}
\label{thm:robustness}
Let $P$ be a weighted combinatorial optimization problem and let $A$ be a deterministic algorithm for it using symbolic perturbation. Then $A$ is \emph{robust}, i.e. the following properties hold for any null case $\mathcal{C}$.
	\begin{itemize}
	\item[a)]  If $A$ returns a solution from $L$ for all problem instances except those in $\mathcal{C}$, then this also holds for the instances in $\mathcal{C}$.
		\item[b)]  If $A$'s runtime (memory consumption) only depends on the taken tree path and is bounded by some function $t:P \to \NN$ for all problem instances except those in $\mathcal{C}$, then this bound also holds for the instances in $\mathcal{C}$.
		\item[c)] If the cost of the solution computed by $A$ is bounded by a function $b:P \times \Q^\NN \to \RR$ that is continuous in  the second parameter, then this bound also holds for the instances from $\mathcal{C}$.
	\end{itemize}
\end{theorem}
From part (c) of the theorem one can derive further properties of algorithms using symbolic perturbation. These properties are established by Lemma~\ref{lemma:opt-continuous}.
\begin{corollary}
Let $P$ be a weighted combinatorial optimization problem and let $A$ be a algorithm using symbolic perturbation. Then, for any null case $\mathcal{C}$, 
\begin{itemize}
\item[d)] if $A$ is a optimal for all inputs except those in $\mathcal{C}$, then $A$ is also optimal for inputs from  $\mathcal{C}$.
\item[e)] if $A$ is a $c$-approximation for all inputs except those in $\mathcal{C}$, where $c:P \times \Q^\NN \to \RR$ is a function that is continuous in  the second parameter, then $A$ is also a $c$-approximation for inputs from  $\mathcal{C}$.
\end{itemize}
\end{corollary}

\tinytitle{Proof of Theorem~\ref{thm:robustness}.} 
Let $\mathcal{C}$ be an arbitrary null case.
For proving part a) and b), imagine that $A$ traverses the tree path to $l$ when processing an instance $(S,w) \in \mathcal{C}$. Due to Lemma~\ref{lemma:mainlemma}, the same tree path is traversed for all weights from some semi-open set $X$ that contains $w$. As $X  \neq \emptyset$, it follows that $X$ has some non-empty open subset, so not every element of $X$ can be a member of $\{x \mid (S,x) \in \mathcal{C}\}$. Therefore, there is an instance $(S,x) \notin \mathcal{C}$ where $A$ behaves like it does in instance $(S,w)$ and thus returns the same solution and has the same resource requirements.

It remains to prove proposition c). We have that 
\[
\sup\left\{ \frac{A(S,w)}{b(S,w)} \mid w \in W, (S,w) \notin \mathcal{C}\right\} \leq 1 \ .
\]
\fullversion{%
 Lemma~\ref{lemma:mainlemma} implies that $A(S,\cdot)$ is locally continuous. As $b(S,\cdot)$ is continuous, we can apply Lemma~\ref{lemma:concatenation} two times for showing that 
 $A(S,w) / b(S,w)$ is locally continuous in $w$.
 }%
 \extendedabstract{%
 Lemma~\ref{lemma:mainlemma} implies that $A(S,\cdot)$ is locally continuous. As $b(S,\cdot)$ is continuous, one more simple lemma (see full paper) shows that 
 $A(S,w) / b(S,w)$ is locally continuous in $w$. }
  Theorem~\ref{thm:math} gives
 \[
 \sup\left\{ \frac{A(S,w)}{b(S,w)} \mid w \in W \right\} = \sup\left\{ \frac{A(S,w)}{b(S,w)} \mid w \in W, (S,w) \notin \mathcal{C}\right\} \leq 1 \ .
 \]
\hbox{}\hfill{$\Box$} 

\section{An example}
\label{sec:example}

This section intends to demonstrate how the concept of symbolic perturbation helps to simplify the analysis of algorithms. As our example we consider the classical Hu-Tucker Algorithm for the Alphabetic Tree Problem. It is one of the ancient examples of a simple problem admitting a simple algorithm for solving it, whereas the correctness proof for the algorithm is highly complicated.

In the Alphabetic Tree Problem we are given a sequence of $n$ letters $b_1,\ldots,b_n$ having certain weights $w_1,\ldots,w_n \in \Q^+_{0}$. The objective is to find a binary tree $D$ whose leaves in left-to-right order are exactly $b_1,\ldots,b_n$ minimizing
\begin{equation}
\label{eq:atp}
\sum_{i=1}^n w_i \cdot \depth(b_i) \ ,
\end{equation}
where $\depth(b_i)$ is the distance between $b_i$ and the root of $D$.

We first show how the Alphabetic Tree Problem fits into the model of combinatorial optimization problems introduced in Section~\ref{sec:model}. Let $P$ denote the Alphabetic Tree Problem. Then $P$ contains a problem structure $S_n = (n,(Q^+_0)^n,L_n,\cost_n)$ for each natural number $n$. In other words, the problem structure is determined by the number of letters, and the domain of the weights is such that each individual weight can be any nonnegative rational number. The solution set $L_n$ contains all binary trees having $n$ leaves, and the cost function assigns the weighted tree depth as defined in Equation~\ref{eq:atp}. It is straightforward to see that the domain of weights is semi-open and that the cost functions are continuous for any fixed solution, so the Alphabetic Tree Problem fits well into our definition of combinatorial optimization problems.

There is a straightforward dynamic programming method solving the problem in time $O(n^3)$. Hu and Tucker~\cite{Hu71} were the first to derive an $O(n \log n)$ time optimal algorithm. Their method, denoted as $H$ in the following, resembles the well-known Huffman coding scheme~\cite{Huffman52}. 

In the first phase, $H$ maintains a sequence of tree nodes, both internal nodes and leaves. Two nodes are combinable if there are only internal nodes between them in the sequence. In each step, $H$ combines the combinable node pair having minimum total weight. Ties are resolved by choosing among the candidate pairs the one consisting of the leftmost possible nodes. This tie-breaking policy is commonly called \emph{alphabetic ordering}. Combining two nodes means to make them the left and right children of a new internal node. When two nodes are combined, they are both removed from the working sequence, the new internal node takes the former position of its left child, and the weight of the new node is the sum of its children's weights. The first phase ends as soon as there is only one node left.

The first phase might produce a tree, say $D^\prime$, that does not satisfy the ordering restriction, see the instance below. In the second phase, $H$ constructs a tree $D$ where the depth of any leaf is the same as in $D^\prime$. The second step is very simple to implement, we do not go into details here.

One of the crucial points is to prove that the second phase always succeeds, i.e. the first phase always produces a tree $D^\prime$ that can be turned into an alphabetic tree. Consider for example the five leaves $b_1,b_2,b_3,b_4,b_5$. $D^\prime$ could combine the pairs $(b_2,b_3)$ and $(b_1,b_4)$ into two internal nodes $b_{23}$ and $b_{14}$. Then it could combine $(b_{14},b_5)$ into $b_{145}$, and finally $(b_{145},b_{23})$. 
There is certainly no alphabetic tree where $b_1,b_2,b_3,b_4,b_5$ have depth $3,2,2,3,2$, respectively. This example demonstrates that tie-breaking matters, because with the wrong tie-breaking policy the tree $D^\prime$ could have been constructed from an instance where $w_1=w_2=w_3=w_4=w_5$.

The second crucial point to prove is that the tree constructed by $H$ is optimal. Up to today, no simple proof for the correctness of the Hu-Tucker Algorithm is known. 
The concept of symbolic perturbation still does not turn the proof into a truly simple one, but at least the consideration of ties, which plays a important role in the proof given by Hu and Tucker, can be set aside.

This is how Algorithm $H$ fits into the model of deterministic algorithms introduced in Section~\ref{sec:model}: the decision which two nodes to combine in the first phase can be done by repeatedly comparing sums of node weights. This is equivalent to comparing the difference of node weight sums with $0$. For example, when $H$ wants to decide which of the pairs $(b_1,b_2),(b_2,b_3)$
has the smaller total weight, it would compare $(b_2 + b_3) - (b_1+b_2)$ with $0$. We assume that the subtrahend is always the node pair having the leftmost left component; if the left components of the pairs are identical, the pair having the leftmost right component is chosen as the subtrahend.
Note that this viewpoint is still valid if the comparisons are made inside some sophisticated data structure (as required for achieving the $O(n \log n)$ runtime).
 Each branching function is a linear combination 
 \begin{equation} 
 \label{eq:H}
 v(w) = \sum_{i=1}^n c_i w_i \ \text{with} \ c_i \in \{-1,0,1\} \ \text{for} \  i=1,\ldots,n \ \text{and} \ (c_1,\ldots,c_n) \neq (0,\ldots,0) 
 \end{equation}
 of weights. Therefore, the branching functions are clearly continuous. It can further be assumed that after Phase 1 the leaves of the decision tree are reached, because the computations in Phase 2 are completely determined by the tree $D^\prime$ computed in Phase 1 and do not further depend on the weights.
 
Now we want to show that the tie-breaking of $H$ can be described as symbolic perturbation. Here only one direction function $h(a) := a \cdot (2^n,2^{n-1},\ldots,2^1)$ works out for each possible configuration of $n$ leaf weights. It is straightforward to observe that $(\Q^+_0)^n$ continues into direction $h$ at each point. It is also not hard to see that the branching functions of $H$ are always increasing or decreasing into direction $r$, depending on the algebraic sign of the first (leftmost) coefficient with $c_i \neq 0$. 

As each internal node takes the position of its left child, the nodes in the sequence maintained by $H$ are always ordered by the position of their leftmost descendant leaf in the initial sequence. Together with the way the subtrahend is chosen among the two pairs when comparing them, it follows that the first nonzero coefficient in Equation~\ref{eq:H} is always -1. In other words, $v$ always decreases into direction $r$ at $w$, and therefore $H$ behaves according to $r$ when choosing the leftmost pair. 

It follows that $H$ employs symbolic perturbation and therefore can be analyzed under the assumption that some arbitrary null case never occurs. For the analysis of $H$, and also for the analyses of many other algorithms, it is the most convenient to assume that all weights are non-zero, ties never occur during the execution, and there is exactly one optimal solution. Those scenarios can be described as the set of roots of a finite family of non-constant linear functions, and it is not hard to observe that such sets are nowhere open.

\section{Proof of existence}
\label{sec:existence}

In this section we shall show that the technique of symbolic perturbation is not limited to a certain subclass of algorithms, but can in principle be used in order to make any algorithm robust. We remark that the result is not constructive: only the existence of a symbolic perturbation tie breaking policy is proven, not the computability.

\begin{theorem}
\label{thm:existence}
	There is a symbolic perturbation tie breaking policy for any deterministic algorithm for a combinatorial optimization problem.
\end{theorem}

The central argument of the proof is provided by the following lemma.

\begin{lemma}
\label{lemma:openpreimage}
Let $X \subseteq \Q^n$ be a non-empty open set and let $f:X \to \Q$ be a continuous function. Then there exists a non-empty open subset $Y \subseteq X$ with $\sgn(f(Y))$ being constantly $0$, $-1$, or $1$.
\end{lemma}

\fullversion
{
\tinytitle{Proof.} If $f(X)$ is constantly $0$, we are done. Otherwise, there is some $x \in X$ with $f(x)\neq 0$, say w.l.o.g. $f(x)> 0$. The continuity of $f$ implies that each element from some $\epsilon$-neighborhood of $x$ has a positive image. Furthermore, $X$ contains some $\epsilon^\prime$-neighborhood of $x$ because it is open. $Y$ being the $\min(\epsilon,\epsilon^\prime)$-neighborhood of $x$ satisfies the desired property.
}

\fullversion{
\tinytitle{Proof of Theorem~\ref{thm:existence}.}
}%
\extendedabstract{
\tinytitle{Proof sketch of Theorem~\ref{thm:existence}.}
}%
 Let $A$ be an algorithm for problem $P$. For each instance $(S,w)$, $S = (n,W,L,\cost) \in P$, we have to prove the existence of a direction function $h$ satisfying the demands from Definition~\ref{def:symbolic_perturbation}. Let $(v_1,\ldots,v_m)$ be an enumeration of all branching functions in the tree traversed by $A$ for input $(S,w)$.

Consider an infinite sequence of open sets $(U_i)_{i \in \NN}$ such that $U_i \in W$ for each $i \in \NN$ and for any $\epsilon>0$ there is some $i_0 \in \NN$ such that $U_i$ is completely contained in the $\epsilon$-neighborhood of $w$ for each $i \geq i_0$. Such a sequence exists because $W$ is semi-open.

\extendedabstract{
We show how to modify that sequence subject to the goal that for each $v_j$, $1 \leq j \leq m$, it holds that $\sgn(v_j(\bigcup_{i=1}^\infty U_i))$ is constantly 0, 1, or -1 and still each member of the sequence is open.

Having constructed the sequence of open sets, we define the direction function $h$ in a pragmatic way: For any $a>0$ we know that there is some member $U_i$ of our sequence that is completely contained in the $a$-neighborhood of $w$. We choose an arbitrary interior point $x_a$ from $U_i$ and assign $h(-a) = h(a) = x_a - w$. Doing this for any value of $a$ and assigning $h(0) = 0$ defines $h$. 
}
\fullversion
{
In the following we modify that sequence subject to the goal that for each $v_j$, $1 \leq j \leq m$, it holds that $\sgn(v_j(\bigcup_{i=1}^\infty U_i))$ is constantly 0, 1, or -1 and still each member of the sequence is open.

Assume that this goal has already been achieved for $v_1,\ldots,v_{j-1}$. Any sequence member $U_i$ is open, so Lemma~\ref{lemma:openpreimage} implies that for any $i \in \NN$ there is a subset $U^\prime_i \subseteq U_i$ such that $\sgn(v_j(U^\prime_i))$ is constantly -1, 0, or 1. Therefore, there has to be a $k \in \{-1,0,1\}$ with the property that for each $i \in \NN$ there is an $i^\prime \geq i$ with $\sgn(v_j(U^\prime_{i^\prime}))=k$.

We now modify the sequence $(U_i)_{i \in \NN}$ in two steps. First, the sequence is replaced with $(U^\prime_i)_{i \in \NN}$. Then, each component $U^\prime_i$ with $\sgn(v_j(U^\prime_i)) \neq k$ is removed. Both steps preserve the desired property for the functions $v_1,\ldots,v_{j-1}$, and the property is established for $v_j$.

Having constructed the sequence of open sets, we define the direction function $h$ in a pragmatic way: For any $a>0$ we know that there is some member $U_i$ of our sequence that is completely contained in the $a$-neighborhood of $w$. We choose an arbitrary interior point $x_a$ from $U_i$ and assign $h(-a) = h(a) = x_a - w$. Doing this for any value of $a$ and assigning $h(0) = 0$ defines $h$. 

The function $h$ is continuous at $0$, because $d(h(\epsilon),h(0)) < \epsilon$ always holds.
By this construction, $W$ continues into direction $h$ at $w$, and each branching function $v_j$ is either decreasing, constant, or increasing into direction $h$ at $w$.
}
\hfill{$\Box$}

\section{Summary, discussion of related and future work}
\label{sec:summary}

In this work we have proposed symbolic perturbation as a proof technique for optimization algorithms. We believe that in many involved analyses it will pay off to first show symbolic perturbation and then assume that some arbitrary null case does not occur. The concept allows theoreticians to concentrate on the interesting points of proofs, not spending too many efforts on the boring corner cases.
\fullversion{%
Typical examples of null cases are:
\begin{itemize}
\item there are ties that the algorithms has to break.
\item some weights are zero.
\item there is more than one optimal solution.
\end{itemize}
}%

Here is the cooking recipe for verifying that a case $\mathcal{C}$ can be excluded from the analysis of algorithm $A$ for problem $P$.
\begin{enumerate}
\setlength{\itemsep}{-2pt}
\item Verify that $P$ fits into the model given in Section~\ref{sec:model}. In particular, verify that the cost functions are continuous in the weights and that the weight domains are semi-open.
\item Verify that $A$ fits into the model given in Section~\ref{sec:model}. In particular, verify that the branching functions are continuous.
\item Find direction functions $h$ with which the requirements in Definition~\ref{def:symbolic_perturbation} are satisfied. When the algorithm employs alphabetic ordering to break ties like algorithm $H$ in Section~\ref{sec:example} does, functions of the type $h(a) = a \cdot (2^n,2^{n-1},\ldots,2^1)$ are often a good choice.
\item Verify that $\mathcal{C}$ is a null case. It suffices to show that any instance in $\mathcal{C}$ can be turned into an instance not in $\mathcal{C}$ by shifting the weight vector by some infinitesimally small amount. The fact that the set of roots of any finite family of non-constant polynomials is nowhere open might also be helpful.
\end{enumerate}

\tinytitle{Generalizations.} We remark that the results from this work cannot be used in general to exclude null cases from the proof that an algorithm terminates for all instances. An algorithm not terminating could be modeled by an infinite path in the tree $T$. For example, the depth $m$ node on that path could ask whether $w - 1/m >0$, where $w \in \Q^+_0$ is some weight. If yes, the algorithm would leave the infinite path. Consequently, it terminates for all instances except the null case $w=0$, no matter how ties are broken.

Another way to model infinite loops is to generalize the decision trees traversed by an algorithm to  finite directed graphs. The theorems in this work generalize to that model, and here also the termination of robust algorithms can be proved disregarding some arbitrary null case. However, algorithms like the one described in the preceding paragraph cannot be modeled using only finitely many branching functions.

\tinytitle{Discussion of related results.}

The technique of symbolic perturbation has been proposed in a number of different contexts. One application is the treatment of degenerate cases in implementations of the Simplex algorithm for solving linear programs. When two or more vertices of the polyhedron defined by the linear constraints coincide, it might happen that (a naive implementation of) Simplex runs into an infinite loop. Explicit symbolic perturbation is one of the methods that have been proposed in order to overcome that problem~\cite{Wolfe63}.

The conceptual difference to our results is that we give a method for recognizing implicit symbolic perturbation rather than proposing a way to explicitly computing such a tie breaking policy. It is though worth mentioning that Simplex can be described using only finitely many branching functions for any problem instance, and therefore any symbolic perturbation method for it guarantees termination.

Another application area of symbolic perturbation are geometric algorithms. Here one distinguishes between problem\--dependent and algorithm\--dependent de\-genera\-cies~\cite{Yap90,Schorn94}. A problem-dependent degeneracy can be described as one in the input, e.g. three points located on the same straight line in the convex hull problem. One could also say that problem-dependent degeneracies are points of discontinuity of the function mapping the input space to the solution space. Algorithm-dependent degeneracies occur when some branching function of the algorithm evaluates to zero.
Among others, Yap~\cite{Yap90} has proposed a method to break ties by symbolic perturbation that can be plugged into any algorithm whose branching functions are polynomials. The method has later been extended to rational functions by Neuhauser~\cite{Neuhauser97}.
However, in a number of other articles it is argued that perturbation is not the silver bullet for all kinds of problems with geometric degeneracies~\cite{Burnikel94,Schorn94}. First, because it incurs extra resource requirements, and for most problems there are cheaper ways to deal with degenerate cases. Second, because problem-dependent degeneracies cannot be handled in an appropriate way.

In the area of optimization problems Lemma~\ref{lemma:opt-continuous} ensures that prob\-lem-dependent degeneracies do not exist. Therefore, optimization algorithms with certain types of branching functions can be made robust using the methods from~\cite{Yap90,Neuhauser97}. It is of course questionable if the runtime of an algorithm should be increased just for the sake of a simpler analysis, but it might still be an option if one is satisfied with an algorithm having any polynomially bounded runtime. In our opinion, however, explicit perturbation in this context unnecessarily moves theory further away from practice, and one should preferably show implicit symbolic perturbation.

\tinytitle{Directions for future work.} The research area of robust algorithms yields a number of interesting open problems. 
Firstly, it is currently not known if results similar to ours can be shown for models of inexact computation. A related question addresses combinatorial problems where the weights can  have only integer values. 
 Secondly, this work has addressed only deterministic algorithms, so can the results be generalized to nondeterminism? Finally, there might be situations where an explicit symbolic perturbation method can be plugged into algorithms without increasing the runtime, which makes it possible to disregard null cases even without having proved symbolic perturbation before. Identifying these situations is another issue for future research.
 
\tinytitle{Acknowledgement.} The author wishes to thank Takeaki Uno for a number of helpful comments that have been the inspiration for a substantial improvement of the results.

\newpage

\thispagestyle{empty}

\end{document}